# Optimum strategy for energy degraders and ionization cooling

Francis J.M. Farley[1]

*Department of Physics, Yale University, Newhaven, USA*

Abstract

Methodology for calculating the profile and emittance of a particle beam as it is slowed down in matter, including the effects of multiple scattering, axial magnetic field and lithium lens. Strategies are determined for minimum final emittance. For ionization cooling, boron carbide has advantages over liquid hydrogen if the beam can be focused onto a small enough area, which may be possible using strong axial fields.

---

[1] *Address for correspondence: 8 Chemin de Saint Pierre, 06620 Le Bar sur Loup, France; phone 0493424512; email farley@dial-up.com*

**1  Introduction**

When a beam passes through matter its profile changes due to two effects; the effect of drift distance following the usual rules for a beam in vacuum, and the effects of multiple scattering.  Assuming that the scattering distribution is Gaussian we derive an equation for the combined effects and show how to minimize the output emittance. Refocusing by an axial magnetic field or lithium lens is included. Sections 2 - 8 are based on an unpublished report by C. Carli and the present author[1].

In particle therapy with protons or light ions coming from a fixed energy accelerator, it is necessary to adjust the range by passing the particles through a slab of matter, known as the degrader.  One should use the best strategy to minimize the loss of beam by transverse scattering.

The results can also be used to study the "ionization cooling" of muon beams, with or without a magnetic field.  We derive equations for cooling in an axial field and in a lithium or beryllium lens.  The analysis is valid if the momentum spread and beam angles are small.  In this case the best degrader is boron carbide; liquid hydrogen appears less useful.  But high intensity muon beams come from pions of various momenta decaying at various distances and may include a wide range of muon momenta; they are intrinsically chaotic and hard to focus. Concentrating such particles onto a degrader with axial magnetic fields is discussed at the end of the paper.

**2  Recapitulation of linear beam transport formalism**

The trajectories of particles in a beam may be defined by the lateral distance $x$ and the direction $\theta$ in a transverse plane relative to a reference track.  If the deviations are small, linear differential equations apply and the motion may be treated by a matrix formalism[2–5]. In the absence of coupling between the horizontal and vertical planes and if the influence of momentum spread is small the equation reduces to

$$\mathbf{X_2} = \begin{pmatrix} x_2 \\ \theta_2 \end{pmatrix} = \begin{pmatrix} R_{11} & R_{12} \\ R_{21} & R_{22} \end{pmatrix} \begin{pmatrix} x_1 \\ \theta_1 \end{pmatrix} = \mathbf{R\,X_1} \qquad (1)$$

where the R-matrix is the transport matrix between points 1 and 2. For a drift space of length $t$ in the absence of magnetic fields the transport matrix is

$$\mathbf{R} = \begin{pmatrix} 1 & t \\ 0 & 1 \end{pmatrix} \qquad (2)$$

The beam profile is assumed to be defined[3] by an ellipse which traces the contour in phase space, for example at one standard deviation from the reference particle, the equation of the ellipse being given by the symmetric Sigma matrix

$$\sigma = \begin{pmatrix} \sigma_{11} & \sigma_{12} \\ \sigma_{12} & \sigma_{22} \end{pmatrix} \quad \text{via the equation}$$

$$\mathbf{X^T}\,\sigma^{-1}\,\mathbf{X} = 1 \qquad (3)$$

where $\mathbf{X^T} = (x \quad \theta)$ is the transpose of the vector $\mathbf{X}$.

When written out this implies

$$\sigma_{22}\,x^2 - 2\,\sigma_{12}\,x\,\theta + \sigma_{11}\,\theta^2 = \Delta = \sigma_{11}\,\sigma_{22} - \sigma_{12}^{\,2} \qquad (4)$$

For an upright ellipse $\sigma_{12} = 0$ and the standard deviations in position and angle are $\sigma_x = \sqrt{\sigma_{11}}$ and $\sigma_\theta = \sqrt{\sigma_{22}}$. $\Delta$ is the determinant of the sigma matrix and the area of the ellipse in phase space is $\pi\sqrt{\Delta}$. It will be convenient to omit the factor $\pi$ and refer to $\varepsilon = \sigma_x \cdot \sigma_\theta = \sqrt{\Delta}$ as the emittance of the beam.

The sigma matrix defining the phase space ellipse is transported[3] according to

$$\sigma_2 = \mathbf{R}\,\sigma_1\,\mathbf{R}^T \tag{5}$$

Applying this formula to a drift space $t$ one obtains

$$\sigma_2 = \begin{pmatrix} 1 & t \\ 0 & 1 \end{pmatrix}\begin{pmatrix} \sigma_{11} & \sigma_{12} \\ \sigma_{12} & \sigma_{22} \end{pmatrix}\begin{pmatrix} 1 & 0 \\ t & 1 \end{pmatrix} = \begin{pmatrix} \sigma_{11} + 2t\sigma_{12} + t^2\sigma_{22} & \sigma_{12} + t\sigma_{22} \\ \sigma_{12} + t\sigma_{22} & \sigma_{22} \end{pmatrix} \tag{6}$$

For small values of $t$ the change in the sigma matrix is

$$d\sigma = \begin{pmatrix} 2\sigma_{12} & \sigma_{22} \\ \sigma_{22} & 0 \end{pmatrix} dt \tag{7}$$

If one integrates (7) doing the off-diagonal terms first, one recovers eqn (6) exactly.

**3  Phase space ellipse and density function**

In a real beam the particles will not occupy exactly the inside of the ellipse defined above, but will be distributed according to some density function, such as the Gaussian

$$\rho(x,\theta) \propto \exp\left\{-\left(\sigma_{22}x^2 - 2\sigma_{12}x\theta + \sigma_{11}\theta^2\right)/2\Delta\right\}$$

(8)

$$= \exp\left\{-\frac{\left(\theta - \frac{\sigma_{12}}{\sigma_{11}}x\right)^2}{2\left(\sigma_{22} - \sigma_{12}^2/\sigma_{11}\right)}\right\} \times \exp\left\{-x^2/2\sigma_{11}\right\}$$

For a given $x$ its distribution in $\theta$ is centred at
$\theta_0 = x\sigma_{12}/\sigma_{11}$ and the variance $\sigma_{22} - \sigma_{12}^2/\sigma_{11}$ is independent of $x$.
If one integrates the first Gaussian over all $\theta$ the result will be an area which is independent of $x$. Then the distribution in $x$ will be given by the second Gaussian with variance $\sigma_{11}$.

By rearranging (8) in a different way one can show that at a fixed $\theta$ the variance of $x$ is $\sigma_{11} - \sigma_{12}^2/\sigma_{22}$, and integrating over all $x$ the variance of $\theta$ is $\sigma_{22}$.

The sigma matrix formalism is valid also for the RMS values in non-Gaussian distributions.

**4  Effect of scattering in the degrader**

The following discussion is limited to linear optics with reasonably small angles in the beam and small momentum spread. It may be a useful guide in the more general case.

If a charged particle of atomic number $Z$ with velocity $v/c = \beta$ and momentum $p$ (in MeV/c), traverses a thin slab of matter of thickness $dt$ and radiation length $X_0$, it suffers many small angle elastic scatterings. This results[6] in a distribution in angle which is approximately Gaussian with variance $V_\theta = Kdt/(p\beta)^2$ where $K = 200Z^2/X_0$. For a thin slab the scattering cannot change $x$ nor the centre of the distribution in $\theta$. Therefore $\sigma_{11}$ and $\sigma_{12}$ are unaffected. The variance in $\theta$ is increased by the amount $V_\theta$, so

$$d\sigma = \begin{pmatrix} 0 & 0 \\ 0 & V_\vartheta \end{pmatrix} \qquad (9)$$

Combining with (7) the overall change in the sigma matrix for a thin slab is

$$d\sigma = \begin{pmatrix} 2\sigma_{12} & \sigma_{22} \\ \sigma_{22} & K/p^2\beta^2 \end{pmatrix} dt \qquad (10)$$

To solve for a thick slab one can integrate term by term, starting with $\sigma_{22}$. Let $t = 0$ at the beginning of the degrader and designate the $\sigma$-components of the initial beam with the additional index $0$. Then the solution for the component $\sigma_{22}$ is

$$\sigma_{22}(t) = \sigma_{22}^0 + C(t)$$

$$\text{with } C(t) = \int_0^t K/(p\beta)^2 \, ds \qquad (11)$$

In the integral one must insert the correct variation of $p\beta$ with distance $s$ in the degrader, obtained from the range energy relation.

Inserting this solution into the equation for the component $\sigma_{12}$ leads to

$$\sigma_{12} = \sigma_{12}^0 + t\,\sigma_{22}^0 + B(t)$$

$$\text{with } B(t) = \int_0^t C(s) \, ds \qquad (12)$$

Using this result to find $\sigma_{11}$ gives

$$\sigma_{11}(t) = \sigma_{11}^0 + 2t\sigma_{12}^0 + t^2\sigma_{22}^0 + A(t)$$

with $A(t) = 2\int_0^t B(s)\,ds$ (13)

At the exit of the degrader of thickness $t$ the final sigma matrix can then be written

$$\sigma^{out} = \sigma^{beam} + \sigma^{degrad} \tag{14}$$

where

$$\sigma^{beam} = \begin{pmatrix} \sigma_{11}^0 + 2t\sigma_{12}^0 + t^2\sigma_{22}^0 & \sigma_{12}^0 + t\sigma_{22}^0 \\ \sigma_{12}^0 + t\sigma_{22}^0 & \sigma_{22}^0 \end{pmatrix} \tag{15}$$

is the matrix for the beam at the end of the degrader allowing for the drift space but with no scattering and

$$\sigma^{degrad} = \begin{pmatrix} A & B \\ B & C \end{pmatrix} \tag{16}$$

The degrader matrix, $\sigma^{degrad}$, contains all the scattering action and is independent of the input beam geometry!  The degrader matrix is the sigma matrix of the beam coming out of the degrader when the input is an ideal pencil beam.  This characterizes the degrader and the determinant gives the emittance $\varepsilon_{degrad} = \sqrt{AC - B^2}$ called the degrader emittance.

## 5 Convolution theorem

A more general input beam can be spit up conceptually into a bundle of pencil beams. The output in this case is the convolution integral of the input beam with the degrader characteristic. Equation (14) shows that the final sigma matrix is the sum of the two component sigma matrices. This is an example of a more general result, the convolution theorem. It can be proved that in all cases the convolution of two elliptical Gaussian distributions represented by matrices $\sigma_1$ and $\sigma_2$ is an elliptical Gaussian distribution represented by the matrix $\sigma = \sigma_1 + \sigma_2$. (The theorem may also be true for some non-Gaussian distributions). It applies also in the presence of an axial magnetic field or lithium lens, which we will discuss below.

## 6 Minimization of output emittance

For a given degrader, the smallest output emittance is obtained by adjusting the input beam to make the determinant of the final sigma matrix $\sigma^{out}$ in (14) as small as possible. We have

$$\det \sigma^{out} = \varepsilon_{beam}^2 + (AC - B^2) + (C\sigma_{11}^{beam} + A\sigma_{22}^{beam} - 2B\sigma_{12}^{beam}) \qquad (17)$$

In this formula the first term, the emittance of the input beam, and the second term, depending only on the degrader, are invariable: but the final term depends on the chosen shape of the input beam. It will be a minimum if

$$C d\sigma_{11}^{beam} + A d\sigma_{22}^{beam} - 2B d\sigma_{12}^{beam} = 0 \qquad (18)$$

while the fixed emittance of the input beam implies

$$\sigma_{22}^{beam} d\sigma_{11}^{beam} + \sigma_{11}^{beam} d\sigma_{22}^{beam} - 2\sigma_{12}^{beam} d\sigma_{12}^{beam} = 0 \qquad (19)$$

Eliminating $d\sigma_{12}$ between these two equations we find that they are satisfied if $\sigma_{11}/\sigma_{12} = A/B$ and $\sigma_{22}/\sigma_{12} = C/B$, implying that the four elements of the beam matrix $\sigma^{beam}$ must be proportional to the corresponding elements of the degrader matrix, so the optimum input beam is given by

$$\sigma^{beam}_{opt} = \frac{\varepsilon_{in}}{\varepsilon_{\deg rad}} \begin{pmatrix} A & B \\ B & C \end{pmatrix} \qquad (20)$$

This is the matrix for the input beam at the end of the degrader (in the absence of scattering). Transposing through distance -d to make an upright ellipse using (6), one finds that the beam should be focused at an image point distant B/C before the end of the degrader; (for thin degraders this corresponds to the centre of the slab).

Inserting (20) into (14) one finds that the output beam apparently diverges from the same image point and has the same shape. The emittance of the output beam comes to

$$\varepsilon^{min}_{out} = \varepsilon_{beam} + \varepsilon_{\deg rad} \qquad (21)$$

<u>This is our main result</u>. With the optimum input beam, the emittance is increased by a constant amount, $\varepsilon_{\deg rad} = \sqrt{AC - B^2}$ characteristic of the degrader.

The shape required for the input beam is determined by the matrix of the degrader. The corresponding standard deviation in angle is $\sigma_\theta = \sqrt{C}$ and the standard deviation in lateral position is $\sigma_x = \varepsilon_{degrad}/\sigma_\theta$. The corresponding quantities for the input beam at the image point should be in the same proportion.

## 7  Numerical Evaluation

For a thin degrader of thickness $t$, $K' = K/(p\beta)^2$ is almost constant, so $C = K't$, $B = K't^2/2$ and $A = K't^3/3$.  The distance from the image point to the end of the degrader is $B/C = t/2$ and the emittance is

$$\varepsilon_{degrad} = K't^2/\sqrt{12} \qquad (22)$$

For a fixed decrement in momentum $\varepsilon_{degrad}$ is inversely proportional to $W^2 X_0$ where $W = -dp/dx$, so the best material will be the one with the largest value of this product.

The ratio of image size to convergence angle, is

$$\beta_\perp = \sigma_x/\sigma_\theta = \varepsilon_{degrad}/C = t/\sqrt{12} \qquad (23)$$

For a thick degrader, when $p\beta$ varies significantly with range, one must calculate the integrals $A$, $B$ and $C$ using the range energy tables.

## 8  Results

As an example the degrader emittance $\varepsilon_{degrad}$ has been calculated for protons slowing down from 250 MeV to 115 MeV in various materials, (the corresponding ranges in water are 38 cm and 10 cm), with the results given in Table 1.

One sees that boron carbide is better than beryllium or graphite, because it combines high density with reasonably low atomic number.  The best material of all is diamond, but it is not usually available in the appropriate sizes.  The relative performance of different degraders is the same for any energy loss and any particle.

In the presence of chromatic aberration with a large momentum spread it may not be possible to achieve the ideal focusing specified by equation (23) in this case the dense degraders may not perform so well.

**9 Axial magnetic field**

In the presence of a magnetic field parallel to the beam the particles will spiral around the lines of force and the transverse spread will be reduced. How much can this improve the performance of the degrader? Solutions to this problem have been given by Farley, Fiorentini and Stocks[7] and Pearce[8]. Here we derive the sigma matrix for a degrader in the axial magnetic field so that the optimum input beam can be specified.

Following references [2] and [3], for particles of momentum $p$ travelling distance $t$ and in axial field $B$, the transverse components of momentum rotate about the field through angle $2kt$ with $2k = eB/\beta\gamma m_0 c^2$. ($k$ is <u>half</u> the wave number at which the particles spiral around the lines of force).

In a frame of reference rotating with angle $kt$, the $x$- and $y$-motions are decoupled to first order and the transport matrix is

$$\mathbf{R} = \begin{pmatrix} \cos(kt) & \sin(kt)/k \\ -k\sin(kt) & \cos(kt) \end{pmatrix} \qquad (24)$$

Applying (5) and letting $t$ tend to zero, one finds

$$\frac{d\sigma}{dt} = \begin{pmatrix} \dot{A} & \dot{B} \\ \dot{B} & \overset{*}{C} \end{pmatrix} = \begin{pmatrix} 2\sigma_{12} & \sigma_{22} - k^2\sigma_{11} \\ \sigma_{22} - k^2\sigma_{11} & -2k^2\sigma_{12} \end{pmatrix} \qquad (25)$$

where the dot indicates differentiation with respect to the distance $t$. Adding the increase of $\sigma_{22}$ due to scattering (as above) the differential equations for the three matrix components are

$$\begin{aligned}\dot{C} &= K' - 2k^2 B \\ \dot{B} &= C - k^2 A \\ \dot{A} &= 2B\end{aligned} \quad (26)$$

For a thin degrader with $K'$ constant this gives

$$\ddot{B} + 4k^2 B = K' \quad (27)$$

The solution is

$$B = \left(K'/4k^2\right)\{1 - \cos(2kt)\} \quad (28)$$

leading directly to

$$\sigma^{\text{degrad}} = \frac{K'}{2}\begin{pmatrix} t\{1-\sin(2kt)/2kt\}/k^2 & \{1-\cos(2kt)\}/2k^2 \\ \{1-\cos(2kt)\}/2k^2 & t\{1+\sin(2kt)/2kt\} \end{pmatrix}. \quad (29)$$

It may be verified that in zero magnetic field ($k = 0$) $A$, $B$ and $C$ are the same as those obtained from (11)-(13) above.

The magnetic field does not change the angle of a track to the axis, so one would expect $\sigma^2_\theta$ to increase uniformly with $t$. This is not the case in the rotating coordinate system. To transform to the non-rotating frame one must add the component $k^2 \sigma^2_x$ so in the laboratory $\sigma^2_\theta = C + k^2 A = K't$ as expected.

The first term in the matrix (29) is $\sigma_{11} = A = \sigma^2_x$. Putting $\phi = 2kt$ gives

$$\sigma^2_x = 2\sigma^2_\theta t^2 (\phi - \sin\phi)/\phi^3 \qquad (30)$$

in agreement with the results of Farley, Fiorentini and Stocks[7] and of Pearce[8] obtained with two quite different methods.

This gives some confidence in (29) which is the degrader matrix in a magnetic field. We see that it becomes an upright ellipse whenever $2kt = 2n\pi$, that is at every complete turn in the field. These are convenient points at which to match the input beam.

Because of the convolution theorem (section 5 above), the procedure for minimizing the emittance of the final beam is the same as before (section 6); the beam shape, calculated at the end of degrader with no scattering, should match the degrader matrix. Then for constant $K'$ the emittance will be increased by

$$\varepsilon_{\deg rad} = (K't/2k)\sqrt{1 - \sin^2(kt)/(kt)^2} \qquad (31)$$

with the second term under the square root becoming negligible after one turn.

At every whole spiral turn in the field this simplifies to $\varepsilon_{degrad} = K't^2/(2kt) = K't^2/2n\pi$ for $n$ spiral turns, compared with $\varepsilon_{degrad} = K't^2/\sqrt{12}$ in zero field. The improvement is a factor *1.814 × n*. One sees from Table 1 that in zero field liquid hydrogen is a factor *5.1* worse than boron carbide; it would need at least *3* turns in the field to make it competitive. This result is valid for thin degraders with any particle and any energy loss.

The sigma matrix for the incoming beam, passing through the magnetic field without scattering, is obtained using (5) with (24).  To simplify, assume that the initial ellipse is upright with $\sigma_{12} = 0$.  Then

$$\sigma^{\text{beam}} = \begin{pmatrix} C^2\sigma_{11} + S^2\sigma_{22}/k^2 & -SC(\sigma_{11}k - \sigma_{22}/k) \\ -SC(\sigma_{11}k - \sigma_{22}/k) & S^2\sigma_{11}k^2 + C^2\sigma_{22} \end{pmatrix} \tag{32}$$

in which $C = \cos(kt)$ and $S = \sin(kt)$

At every whole turn in the field ($kt = n\pi$) the ellipse again becomes upright and one can match to the upright degrader ellipse (29) by adjusting the input values of $\sigma_{11}$ and $\sigma_{22}$.

From (29) the beta value should be $\beta_\perp = \sigma_x/\sigma_\theta = 1/k = \lambda/\pi$

where $\lambda$ is the spiral wavelength in the solenoid.

**10 Lithium lens**

If the beam is traveling through a rod of radius $a$ carrying a current uniformly spread thorugh its cross section, the magnetic field $B$ inside the rod is proportional to the distance $x$ from the axis, $B = B_0 x/a$, where $B_0$ is the field at the surface of the rod.  The field is everywhere perpendicular to the track of the particle and the bending in distance $dt$ is

$$\frac{d\theta}{dt} = -\frac{eB}{\beta\gamma m_0 c^2} = -\frac{eB_0 x}{\beta\gamma m_0 c^2 a} = -k^2 x \tag{33}$$

with $\quad k = \sqrt{\dfrac{eB_0}{\beta\gamma m_0 c^2 a}} \tag{34}$

(In this case the wave number at which the particles oscillate in the focusing system is $k$).

The transport matrix is (24) with the new value of $k$ and applying (5) one finds again equation (25). Therefore the differential equations (26) apply, the solution is again (29) and the conclusions of Section 9 apply.

The diffusion of particles in a lithium lens was treated by Fernow and Gallardo[9] by a different method; their distribution ellipse is specified by the parameters F, G and H given in their equation (12) which are identical to our A, C and B in (29) above with the ansatz $\omega = k$, and $\theta_c^2 = K'$.

From Table 1 we see that in zero field lithium is 3.03 times worse than boron carbide. With the new definition of $k$, for $n$ complete oscillations in the field the improvement factor is *3.628 × n* so after one or more oscillations a lithium lens degrader will be better than boron carbide in zero field.

**11 Ionization cooling**

Ionization cooling is proposed for reducing the emittance of muon beams before acceleration[10, 11]. If one reduces the forward momentum $p$ by the amount $\delta p$ in an energy degrader of thickness $t$ with no scattering the transverse momenta would be reduced in the same proportion; then when one restores the longitudinal momentum by acceleration the original emittance $\varepsilon$ would be reduced by

$$\delta \varepsilon_a \;=\; \varepsilon \, \delta p / p \;=\; -\,\varepsilon \, W t / p \qquad (35)$$

where $W = -dp/dx$.

However the scattering in the degrader will in the optimum case increase the emittance by (22). This exceeds (35) if the degrader thickness $t$ is greater than $t_m$ given by

$$t_m = \frac{\sqrt{12}}{K'} \frac{\varepsilon W}{p} \qquad (36)$$

If we are to have useful cooling $t$ must be less than $t_m$. Putting $\eta = t/t_m$ one finds that the net cooling is

$$-\delta\varepsilon_{net} = \frac{\varepsilon W}{p} \eta (1-\eta) t_m \qquad (37)$$

which is a maximum when $\eta = 0.5$. The maximum cooling for a single degrader is then

$$-\left(\frac{\delta\varepsilon}{\varepsilon}\right)_{opt} = \frac{W}{2p} t_{opt} = \frac{\varepsilon \sqrt{3} \, W^2 X_0}{400} \qquad (38)$$

with the optimum thickness

$$t_{opt} = \frac{\varepsilon \sqrt{3} \, W \, p \, X_0}{200} \qquad (39)$$

Equation (38) shows that the fractional reduction in emittance is independent of beam momentum. It is proportional to $\varepsilon$ and so becomes smaller and smaller as the beam is cooled. Getting below *50 mm.mR* looks difficult, see Table 2.

To get efficient cooling it would be good to work at low energy where $W = dp/dx$ is large. However this is not possible because the spread in energy will increase too much[9, 10]. This factor forces us to work close to minimum ionization at $\beta\gamma = 3$. As an example, for *0.315 GeV/c* muons the optimum cooling in a single stage and the corresponding energy loss $\Delta E$ are given in Table 2 for two degrader materials, boron carbide and liquid hydrogen, with various initial emittances.

With initial emittance $\varepsilon = 1000$ mm.mR, for example, five degraders of liquid hydrogen, total length *320 cm*, interleaved with focusing systems would reduce the emittance by *14.6%* with an energy loss, to be replaced by acceleration, of *91.5 MeV*. In contrast one layer of boron carbide *22 cm* long with an energy loss of *94 MeV* would reduce the emittance by *15.6%*.

Liquid hydrogen, because of its low value of $W^2 X_0$ (see Table 1) does not appear attractive. We saw in section 9 that an axial magnetic field can improve the performance, but it needs three or more spiral turns of the beam inside the hydrogen to be competitive. This implies a field of order *10 T*.

If the degrader is a lithium lens (Section 10) then one complete oscillation of the beam inside the lithium is sufficient to make it marginally better than boron carbide.

All this presupposes that in each case the beam is focused onto the degrader in the optimum way to achieve minimum increase in emittance as specified in sections 6 and 9 above. The need to use a high density degrader and focus to a small spot was already emphasized by Neuffer[11]. Boron carbide is superior to liquid hydrogen provided that the beam can be focused to the small values of $\beta_\perp$ implied by (23). For chaotic muon beams with large $\Delta p/p$ this may not be possible with conventional systems. Concentrating the beam with axial solenoid fields is discussed below.

**12 Equilibrium emittance in an axial field**

If the beam makes several turns in a solenoid field $B$, equation (31) shows that $\delta\varepsilon_{degrad}$ increases as $t$ rather than $t^2$, so in this case there is no optimum thickness. Comparing (31) with (35) and using the definition of $K'$ there will be cooling if

$$\varepsilon \;>\; \varepsilon_{equilib} \;=\; \frac{200}{2\,k\,p\,\beta^2\,W\,X_0} \tag{40}$$

This determines the equilibrium emittance below which no cooling will occur.

In a field of $1\ T$ the muon angular cyclotron frequency $(eB/m_0c)$ is $8.506 \times 10^8\ s^{-1}$ so, using the definition of $k$ (section 9),

$$\varepsilon_{equilib} \;=\; \frac{200\,c}{\beta^2\,W\,X_0\,(eB/m_0c)\,m_0c} \;=\; \frac{66.8}{\beta^2\,W\,X_0\,B(Tesla)} \tag{41}$$

For $\beta \sim 1$ this is independent of the muon momentum.

In liquid hydrogen in a typical field of $7\ T$, $\varepsilon_{equilib}$ comes to $386\ mm.mR$. The focusing wavelength comes to

$$\lambda \;=\; \pi/k \;=\; \frac{2\pi\beta\gamma\,c}{eB/m_0c} \;=\; 222\,\beta\gamma\,/\,B(Tesla)\quad cm \tag{42}$$

For example, with $\gamma = 3$ in a field of $7\ T$, $\lambda = 94\ cm$ and we have seen above that in hydrogen one needs 3 turns in the field to be competitive. This means that each section should be at least 2.8 m long, implying an energy loss of 80 MeV. For smaller steps in energy, boron carbide without field is superior.

However, of the substances in Table 1, liquid hydrogen has the highest value of $WX_0$ and therefore the smallest equilibrium emittance in a given axial field.

### 13. Equilibrium emittance in a lithium lens

In a lithium lens equation (31) applies with $k$ given by (34). Then

$$\varepsilon_{equilib} = \frac{100}{\beta^2 m_0 c W X_0} \sqrt{\frac{c\,a}{\beta \gamma (e B_0 / m_0 c)}} = 416 \sqrt{\frac{a(cm)}{\beta^5 \gamma B_0(T)}} \quad mm.mR \qquad (43)$$

while the focusing wavelength is

$$\lambda = 2\pi/k = 37.3 \sqrt{\frac{\beta \gamma\, a(cm)}{B_0(T)}} \quad cm \qquad (44)$$

For example a lithium rod of radius *a = 2 cm,* carrying a current of *500 kA* would have a surface field *B₀ = 5 T*. In this case $\varepsilon_{equilib}$ = *152 mm.mR* and the wavelength of oscillations is *41 cm*. The energy loss in this length would be *36 MeV*. For smaller steps in energy boron carbide in zero field would be better.

Beryllium with no field is 2.6 times better than lithium (see Table 1) and has been considered for a lens[12]. For the same diameter and current as above, $\varepsilon_{equilib}$ is slightly worse at *200 mm.mR,* the focusing wavelength is the same (*41 cm*), but the energy loss in this distance is *120 MeV,* so the cooling effect is much greater. For shorter lengths the lens action becomes progressively less significant.

### 15. Focusing with solenoid fields

High intensity muon beams come from pions of various momenta, decaying at various distances, and may include a wide range of muon momenta[13]. Therefore they are "chaotic" and cannot be brought to a precise focus. They can nevertheless be squeezed to small values of $\beta_\perp$ by means of a solenoid with gradually increasing magnetic field. Using the (well-known) equations for a particle of momentum $p$ spiraling at angle $\phi$ to an axial field $B$, we compute the available phase space. The particle wraps itself round a cylinder of radius $r$ with

$$p \sin\phi = QBr \tag{45}$$

where $Q$ is the constant of proportionality.

If $B$ is increasing with axial distance $z$ the radial field $B_r$ at radius $r$ is obtained using Gauss' theorem

$$B_r = -\frac{r}{2}\frac{dB}{dz} \tag{46}$$

In distance $ds = dz/cos(\phi)$ along the orbit this radial field changes the spiral angle $\phi$ by

$$d\phi = \frac{QB_r ds}{p} = \frac{\sin\phi}{\cos\phi}\frac{B_r dz}{Br} = \frac{\sin\phi}{\cos\phi}\frac{dB}{2B} \tag{47}$$

so

$$\frac{d(\sin\phi)}{\sin\phi} = \frac{dB}{2B} \tag{48}$$

implying

$$\sin^2\phi = B/B_0 \tag{49}$$

where $B_0$ is the field at which $\phi$ becomes $90^o$ and the spiraling particle is reflected (magnetic mirror). Substituting $sin(\phi)$ from (45) one finds that $Br^2$ is constant, implying that the flux through the orbit is a constant of the motion. Moving into stronger or weaker fields, the particles continue to spiral around the same lines of force. For small spiral angles

$$\phi^2 \;=\; B/B_0 \;=\; (r_0/r)^2 \tag{50}$$

For a particular particle, $B_0$ and $r_0$ are constant but depend on the starting conditions; $\phi r$ is invariant.

Now consider a solenoid of radius *R*. For a particle starting at radius *b* from the axis of the solenoid, the maximum spiral radius for which the orbit cannot hit the walls is *(R-b)/2*. This estimate is correct when *b=0* and when *b=R* but underestimates the available orbits at intermediate values of *b*; so our estimate of the available phase space will be conservative. The corresponding maximum spiral angle is $\phi = QB(R-b)/2p$ and the solid angle available for particles in the annulus *b* to *(b+db)* is $\pi\phi^2$, making the corresponding 4-dimensional phase space

$$d\varepsilon_4 \;=\; 2\pi^2 b.db \left(\frac{QB}{2p}\right)^2 (R-b)^2 \tag{51}$$

Integrating over *b* from *0* to *R* gives $\varepsilon_4$ and we take the square root to get the corresponding 2-dimensional phase space $\varepsilon_2$ with the factor $\pi$ omitted following the convention used throughout this paper,

$$\varepsilon_2 \;=\; \frac{QBR^2}{p\sqrt{24}} \;=\; \sqrt{\frac{3}{8}}\,\frac{BR^2}{p} \tag{52}$$

in which we have set *Q=3* corresponding to the convenient choice of units *B* in *T*, *p* in *MeV/c* and *r* in *cm.*

We see that if the field is increased progressively to squeeze the orbits as per eqn(50), keeping the flux through the solenoid constant, the phase space available for the particles remains unchanged. (This will break down when ϕ becomes larger than about *0.5* radians; sin(ϕ) then increases more slowly than ϕ, our approximation is no longer valid and the particles will eventually be reflected; but we are not straying into this region).

Using eqn(45)

$$\beta_\perp = r/\phi = p/QB = p/3B \qquad (53)$$

Putting in numbers, for muons of momentum *200 MeV/c* squeezed by a solenoid of radius *9 cm* with field *10 T*, the available normalized 2-dimensional phase space is *67 π mm.R*, which would comfortably accept the beams of *20 π mm.R* requiring to be cooled[14]. The corresponding value of $\beta_\perp$ is *7 cm* so these muons would be matched rather well into a boron carbide degrader. After the degrader the beam could be expanded for the acceleration stage by reducing the field, following eqn(50): for example with the field reduced to *1 T* the solenoid radius would be increased to *28 cm*. The high field regions can be quite short.

## 14. Summary

For proton therapy with a fixed energy accelerator the best degraders are boron carbide, beryllium and graphite. To minimize the final emittance it is essential to match the beam to the degrader using equation (23).

For ionization cooling the beam should again be matched to the degrader. Liquid hydrogen may not be competitive even in a strong axial field. A lithium lens (or beryllium lens) is more promising, but none of these can cool below $\varepsilon_{equilib}$ given by equations (41) and (43), typically in the range *150 — 400 mm.mR*. In contrast there is no theoretical limit to the cooling by a low *Z*, dense degrader such as boron carbide; it just becomes very slow below *50 mm.mR* so the muon lifetime will be the critical factor. The problem of focusing intense muon beams to low values of $\beta_\perp$ may be alleviated by using high axial fields.

**Acknowledgements**

These studies were initiated when the author worked with Pierre Mandrillon on the European Light Ion Medical Accelerator (EULIMA) which was the forerunner of current cancer treatments with beams of carbon. I thank him for his support and encouragement. Christian Carli [1] developed the methodology used in Section 4.

I am grateful to Robert B. Palmer, R.C. Fernow and D.M. Kaplan for stimulating discussions.

Table 1

Degrading a proton beam in various materials from 250 MeV to 115 MeV (range 38 cm and 10 cm in water).

| Degrader | Density | Energy loss W at minimum | Radn length $X_0$ | $\varepsilon_{degrad}$ | $W^2 X_0$ | $W X_0$ |
|---|---|---|---|---|---|---|
| | g/cm$^2$ | MeV/cm | cm | mm.mR | MeV$^2$/cm | MeV |
| water | 1.000 | 1.991 | 36 | 117 | 143 | 72 |
| liquid $H_2$ | 0.071 | 0.286 | 865 | 238 | 71 | 247 |
| lithium | 0.534 | 0.875 | 155 | 142 | 119 | 136 |
| beryllium | 1.848 | 2.946 | 35 | 55 | 306 | 103 |
| boron carbide | 2.520 | 4.256 | 20 | 47 | 361 | 85 |
| graphite | 2.265 | 3.952 | 19 | 57 | 294 | 75 |
| diamond | 3.510 | 6.125 | 12 | 37 | 450 | 75 |

$\varepsilon_{degrad}$ is the emittance added to the beam with optimum focusing. The smallest degrader emittance $\varepsilon_{degrad}$ is the best.

The penultimate column gives the value of $W^2 X_0$ for each material, the largest value giving the best result. The values of $W X_0$ (final column) are used in sections 12 and 13.

Table 2

Optimum cooling and corresponding energy loss for *0.315 GeV/c* muons with a single degrader of boron carbide or liquid hydrogen for various initial emittances.

| Degrader material | Initial emittance $\varepsilon$ (mm.mR) | Optimum thickness $t_{opt}$ (cm) | Decrease in emittance $\delta\varepsilon/\varepsilon$ (%) | Energy loss $\Delta E$ (MeV) |
|---|---|---|---|---|
| $B_4C$ | 1000 | 22 | 15.6 | 94 |
|  | 100 | 2.2 | 1.56 | 9.4 |
|  | 30 | 0.7 | 0.46 | 3.0 |
| $H_2$ liquid | 1000 | 64 | 3.1 | 18.3 |
|  | 100 | 6.4 | 0.31 | 1.8 |
|  | 30 | 1.9 | 0.09 | 0.55 |